\documentclass{article}
\usepackage[utf8]{inputenc}
\usepackage{xcolor}
\usepackage{authblk}

\title{Atmospheric Modelling and Retrieval} 
\author[1]{Jonathan J. Fortney}
\author[2]{Joanna K. Barstow}
\author[3]{Nikku Madhusudhan}
\affil[1]{Department of Astronomy and Astrophysics, University of California, Santa Cruz}
\affil[2]{School of Physical Sciences, The Open University, UK} 
\affil[3]{Institute of Astronomy, Cambridge University}
\date{}
\usepackage{natbib}
\usepackage{graphicx}
\usepackage{geometry}
\usepackage[utf8]{inputenc}

\newcommand{\apj}{ApJ}
\newcommand{\apjl}{ApJ~Lett.}

\newcommand{\icarus}{Icar.}
\newcommand{\nat}{Natur.}
\newcommand{\pasp}{PASP}
\newcommand{\aj}{AJ}
\newcommand{\mnras}{MNRAS}
\newcommand{\apjs}{ApJ~Supp.}
\newcommand{\aap}{A\&A}
\newcommand{\jqsrt}{Journal of Quant Spect and Rad Tran}
\newcommand{\baas}{Bulletin of the AAS}
\newcommand{\aapr}{Astronomy and Astrophysics Review}

\begin{document}
\maketitle

\section{Introduction} 

We seek to learn about the atmospheres of exoplanets both to understand their  physical and chemical processes at work, as well as to potentially tie atmospheric abundances to theories of planet formation and evolution.  Such work relies on spectroscopic investigation.  The interpretation of spectra of exoplanetary atmospheres is heavily dependent on atmospheric models. The accuracy and reliability of these models is especially critical in the absence of \emph{in situ} measurements that in the Solar System provide us with the ``ground truth'' to anchor remote sensing observations.

\begin{figure}[t]
\includegraphics[width=1.0\textwidth]{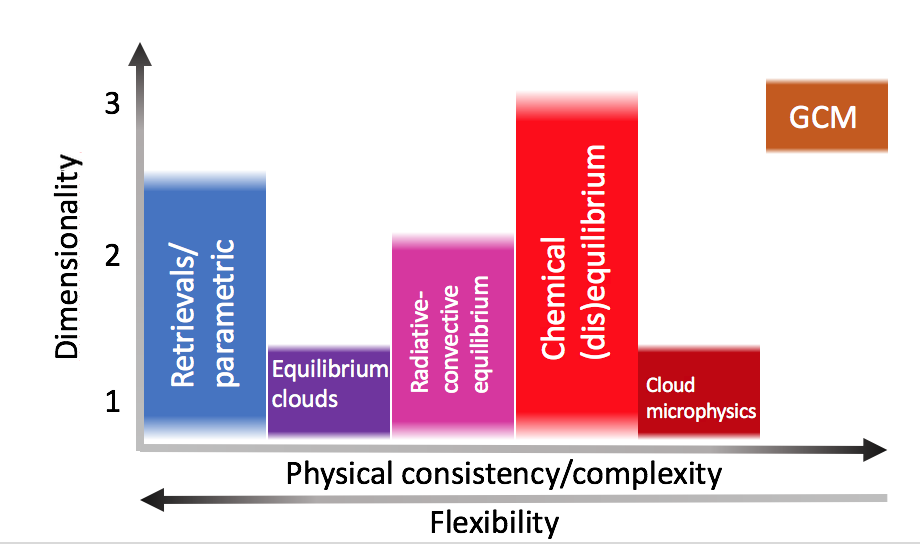}
\caption{This schematic illustrates the typical complexity/flexibility and dimensionality of various types of exoplanet atmospheric modelling tool. Retrieval models can contain varying degrees of physical constraint. Equilibrium clouds includes models such as \cite{Ackerman2001}; microphysical clouds refers to models used by e.g. \cite{Gao:2020aa}.}
\label{fig:fig1}
\end{figure}

Atmospheric models for exoplanets vary in type and complexity (Figure~\ref{fig:fig1}). Models attempt to understand the state of an atmosphere, including the temperature structure, atmospheric abundances, and cloud composition and distribution.  ``Self-consistent'' models, where a wide range of atmospheric physics and chemistry are specified, are typically either 1D and iterate to radiative-convective equilibrium, \citep[e.g.][]{Fortney08a}, or 3D, general circulation models (GCMs), which use the equations of fluid dynamics (or some simplifications thereof) to describe the atmospheric dynamics, \citep[e.g.][]{Showman09,Heng15}.  Either in 1D or 3D, such models need to include sub-models that treat chemical abundances as a function of pressure and temperature \citep{Lodders02}, the chemical effects of incoming stellar UV flux and vertical mixing \citep{Moses11}, as well as cloud microphysics and opacities \citep{AM01, Helling14}.  With all physical effects either explicitly treated within some framework (or explicitly ignored) one calculates a solution for a given atmosphere's state, and may then generate an emission, reflection, or transmission spectrum, which is then compared to observations. 

By contrast, parametric models are normally incorporated into a data-driven ``retrieval'' approach \citep{Mad18x}. A variety of atmospheric parameters that control, for instance, the temperature structure, atmospheric abundances, or a small number of cloud parameters, can be tuned in order to provide the best match to a spectrum, typically with minimal physical assumptions.  Within this framework one performs a wide exploration of a range of possible atmosphere models that can yield a best-fit to an observed spectrum.  Such models can find solutions outside the confines of self-consistent models.  They are most important in providing a sound statistical assessment of retrieved parameters, including mixing ratios of atmospheric gases and the shape of the atmospheric temperature profile.  These models have mostly been developed in the 1D context, but recent work has shown the importance of 2D and 3D frameworks for inhomogeneous atmospheres \cite{Caldas19,MacDonald20,Taylor20}. 

These families of different model types represent a compromise between computation speed/flexibility and physical accuracy; as such they have different uses. For example, highly physics-based GCMs are typically used to simulate detailed conditions on a specific planet, with a pre-determined set of inputs.  1D self-consistent models can be calculated across large ``grids" of parameter space (as is often done in classic stellar atmospheres modeling) to investigate the role of surface gravity, metallicity, and intrinsic fluxes \citep{Burrows97,Fortney08a,Molliere15,Gandhi17}.  Retrievals can be run on individual planets, or larger samples of spectra for many planets, to search for trends \citep{Barstow17,Line17,Welbanks19}, which can be compared to expectations from self-consistent models.


Atmospheric modeling and retrieval methods ultimately seek to answer central questions regarding the diversity of physical and chemical processes in exoplanetary atmospheres and their possible connections to planetary formation conditions. Some of the major questions include:

\begin{itemize}
 \item How do atmospheric properties and processes in exoplanets depend on the macroscopic planetary conditions, such as the irradiation, elemental composition, bulk properties (masses and radii), and stellar environment? 
 
 \item How do different atmospheric processes (e.g., chemistry, energy transport, dynamics and escape) interact with each other across the diversity of exoplanets? 
 
 \item How representative are atmospheric properties of exoplanets of their interior conditions? For giant planets, are chemical abundances in the observable atmosphere representative of the entire H/He envelope? For low-mass planets what can observable compositions tell us about atmosphere-surface interactions? 
    
 \item How well can we extract information about 3D atmospheric properties of exoplanets from their 1-D disc- and terminator averaged observations? 

 \item What are the limits on accuracy and precision for exoplanet spectroscopy, in emission, transmission and reflection?

\end{itemize}

\section{State of the art}  

Exoplanet atmosphere modeling emerged as an offshoot both of solar system planetary atmosphere modeling \citep{Burrows97,Gouk00} as well as stellar atmosphere modeling \citep{SS98,Barman01}.  As such, a very wide variety of methods, with heritages, assumptions, strengths, and pitfalls, from a diverse array of modeling frameworks have long come into play, and continue to do so.  This kind of merging of cultures and disciplines can be seen in many aspects of modeling.  An example is the technique, ``correlated-k,'' of tabulating strongly non-grey molecular opacity with hundreds of millions of individual lines \citep{Goody89}, commonplace for decades in the solar system, which has gradually been adopted by many exoplanet atmosphere codes \citep{Amundsen17} (but not without healthy pushback and discussion).

Pioneering methods of retrieval, to quantify the quality of model fits and robustly assess uncertainties on atmospheric abundances and temperature structures, were independently developed in an exoplanet astrophysics context \citep{Madhusudhan:2009,Madhu11} alongside modification and updating of methodologies created in Earth and solar system remote sensing \citep{Rodgers00,Irwin08}.  The inherent similarities in these methods was investigated \citep{Line13}, and there is now considerable work to understand and quantify the role of different modeling choices in affecting accuracy and precision on constraints \citep{Barstow20}.

While retrieval codes were originally created as 1D ``average" model atmospheres, to compare to both transmission and emission spectra, there has been significant thought in the past five years about limitations of this 1D framework, in particular for hot Jupiters which have strongly inhomogeneous atmospheres from day to night.  Complications could arise from partly cloudy atmospheres \citep{Line16}, temperature inhomogeneities \citep{Feng16,Blecic17,MacDonald20,Taylor20}, day-to-night temperature differences \citep{Caldas19,Lacy20}.  2D and 2D+ retrieval scenarios are now being investigated for interpreting thermal emission phase curves \citep{Irwin20,Feng20}.

Three-dimensional GCM simulations of exoplanet atmospheres mostly have a heritage in Earth and solar system atmosphere modeling, both for giant planets \citep{Showman02,Showman09,Rauscher:2010,Mayne14} and for terrestrial worlds \citep{Joshi97,Leconte13,Carone14}.  However, there is also work underway to create new general GCM tools specifically for exoplanets \citep{Deitrick20}.  Bringing realistic radiative transfer to these simulations is mostly now a solved problem \citep{Showman09,Amundsen:2014}.  However, significant work is currently ongoing on treatments of cloud particle formation and transport via a wide variety of schemes \citep{Lee:2016,Lines:2018,lines:2019,Roman:2019aa}, as well as explorations of simplified chemical networks that can be used within GCMs \citep{Drummond:2018aa,Zhang18b,Venot20}.

\section{Important questions/goals} 

\begin{itemize}

\item \textbf{What is the impact of photochemistry on exoplanet temperature structures and spectra?}  The solar system's giant planets are incredibly cold, leading to a dearth of gases that can readily be photolized to drive further chemistry (e.g., CH$_4$).  For warmer planets, a much richer variety of gases will be found, which may drive extremely complex pathways of photochemical reactions, and create novel pathways for haze formation, that are not seen in the solar system \citep{Zahnle09, Horst18, Chao20}. For terrestrial planets, which unlike giant planets, do not have a vast reservoir of a deep atmosphere in thermochemical equilibrium, photochemistry takes on an ever greater importance, likely with a very wide phase space of possible atmospheric compositions.  It is possible, perhaps even likely, that a wide variety of necessary rate constants and dissociation cross-sections have not yet been determined in the laboratory \citep{Fortney19}.  

\item \textbf{What will we be able to learn about secondary atmospheres from \emph{JWST}?} Secondary atmospheres --- generated by outgassing from a rocky planet --- are more challenging to observe in transmission than primary, H$_2$-He dominated atmospheres due to their higher mean molecular weight resulting in a reduced scale height. Molecular absorption signatures in these atmospheres will be relatively low amplitude. However, with sufficient coadded observations it has been demonstrated \citep{Barstow16,Morley17b} that \textit{JWST} would be capable of detecting various molecules in the atmospheres of favourable targets, such as the TRAPPIST-1 planets. Due to the small signals, likely challenges will be distinguishing atmospheric signatures from those of stellar activity \citep{Rackham18a}, or identifying muted features in the presence of clouds. 

\item \textbf{How are atmospheric abundances impacted by surface/atmosphere interactions for thin primary atmospheres and for secondary atmospheres?}  The role of surface/atmosphere exchange is very uncertain for the atmospheres that will be characterized in the coming decade.  The high temperature and surface pressures are mostly outside the realm of previous work in the solar system.  For instance, the role of a long-lived molten surface altering the abundances in thin H/He dominated atmospheres is just now beginning to be investigated \citep{Kite20}.

\item \textbf{What is the appropriate treatment for cloud and haze in retrieval models, and what can we learn from microphysics simulations?} Clouds and hazes are complex atmospheric phenomena which are challenging to represent, especially in parametric retrieval models which aim to be as computationally efficient as possible. Recent work \citep{mai19,barstow20b} has been done to explore the effects of different parameterisations on retrievals of both simulated and real spectra. As yet, only very basic properties of clouds can be directly inferred from observations, and as a consequence assumptions about characteristics such as composition can only come from microphysical simulations \citep[e.g.,][]{Helling:2016,Helling17,Gao:2020aa}. Symbiosis between such simulations and retrievals will be necessary as we enter the \textit{JWST} era, such that detailed models are used to benchmark parameterisations, and information gained from retrievals can provide observational constraints that allow microphysical models to be refined. 

\item \textbf{For what data sets is the use of a 1D retrieval framework sufficient for determining atmospheric properties?} Recent work has demonstrated that using 1D retrievals, with the inherent assumption of atmospheric homogeneity, may introduce bias when they are applied to planets with significant variations in temperature and cloud coverage around or across the terminator \citep{Line2016,MacDonald20,Lacy20} or the disk-averaged emitting hemisphere \citep{Feng16}. The sensitivity to inhomogeneity and degree of bias, and therefore the necessity of including more complex parameterisations, is dictated both by the wavelength coverage and signal to noise of the observation. Tests on simulated observations with different signal to noise properties, such as those conducted by \cite{Lacy20}, will be necessary to determine this.

\item \textbf{How do we build 2D/3D frameworks capable of extracting information from transit/disc averaged spectra?} Knowledge gained from phase curve observations of hot Jupiters \citep[e.g.,][]{Stevenson:2014,Wong:2019aa,Kreidberg18} tells us how we might expect conditions to vary across the dayside and terminator regions of tidally locked, highly irradiated exoplanets. Whilst spectroscopic observations are available for only a handful of these objects, 3D GCMs provide an alternative source of information about expected atmospheric inhomogeneity. Using existing, detailed observations and simulated outputs from GCMs will allow us to benchmark ways of parameterising variations in temperature, cloud coverage and chemistry. 

\item \textbf{What are \emph{unknown} unknowns, beyond known unknowns?}  So far, we know so little about exoplanetary atmospheres that it is natural to wonder, up until now, if we are even asking the right questions.  It is quite plausible that the first high-precision broad-wavelength-coverage spectra obtained with \emph{JWST} may highlight important issues of atmospheric physics and chemistry that have not yet been highlighted in previous work.

\end{itemize}

\section{Opportunities} 
\begin{itemize}
\item Scheduled for launch in late 2021, \emph{JWST} will enable breakthrough transiting planet spectroscopy from 0.6 to 5 $\mu$m via several instruments, and longward of 5 $\mu$m with the MIRI instrument.  The high signal-to-noise observations across a very broad wavelength range will be transformational for the field, across the range of planets from gas giants orbiting Sunlike stars \citep{Beichman14,Greene16,Lacy20}, to Earth-size planets in orbit around M dwarfs \citep{Barstow16,Morley17b,Koll19}.
\item Scheduled for launch in 2029, Atmospheric Remote-sensing Infrared Exoplanet Large-survey (\emph{Ariel}), is the fourth medium-class mission of the European Space Agency.  Using photometry from 0.4 to 2 $\mu$m and spectroscopy from 2-8 $\mu$m, \emph{Ariel}  will focus on transmission spectra, emission spectra, and phase curves of $\sim$1000 transiting planets across a wide range of planet masses \citep{Tinetti18}.  \emph{Ariel} will provide a statistical assessment of exoplanet atmospheres, to complement the detailed view of a smaller number of targets investigated by \emph{JWST} \citep{Zellem:2019aa,changeat20}.
\item As multiple-node machines become increasingly cost effective, retrievals in 2D or even 3D will become more computationally viable. Optimising codes to run on GPU nodes can also improve efficiency for retrievals (e.g. \cite{kitzmann20}.)

\item A decade after the first detection of an exoplanet atmosphere via high spectral resolution spectroscopy \citep{Snellen:2010}, the field is now coming into its own \citep{Birkby18}. A bounty of high spectral resolution ($R > 25,000$ platforms are just now coming online that will enable a broad array of atmospheric characterization in the optical and near-infrared.  In addition to detecting atoms and molecules, wind speeds \citep{Snellen:2010,Louden:2015} and planetary rotation \citep{Brogi:2015} can be characterized.  The development of spectrographs with a broad simultaneous wavelength coverage will be especially powerful. 

\item Comparative exoplanetary spectroscopy between transiting planets and directly imaged planets.  While the number of directly imaged planets with measured emission spectra is relatively low, the quality of these observations has been dramatically increasing \citep[e.g.,][]{Gravity20}.  In the Extremely Large Telescope (ELT) era, spectra of $\sim$100 directly imaged gas giants, all with known dynamical masses, will be possible \citep{Brandt19}.  Contrasting the atmospheric abundances for these planets on multi-AU orbits, compared to close-in transiting planets, may prove critical in unraveling how atmospheric abundances are tied to planetary formation location \citep{Oberg11,Madhu14,Mordasini16}.

\item As exoplanet observations reveal more detail about their atmospheric structure, techniques and findings from solar system research will become increasingly applicable. Over the last several decades, observations of solar system planets have improved and expanded from observations taken from the ground to detailed imagery from orbiting satellites, and even in some cases in situ measurements using descent probes. As observational constraints have improved, solar system planet models have been forced to develop in complexity and detail. With the exoplanet community facing a similar dramatic change in observational data, we can benefit from the experience of solar system researchers during this process. 
\end{itemize}

\section{Challenges}  
\begin{itemize}
\item Inference from exoplanet observations is hampered by a lack of ground truth. In the solar system, interpretation of remote sensing observations is often anchored by a direct, in-situ measurement of conditions at some time or place on the planet (for example, directly measured temperature and cloud profiles were obtained by the \emph{Pioneer} Venus mission, which has allowed informative priors to be applied for subsequent analyses of the Venusian atmosphere). Similar constraints could be obtained from GCMs or chemical models for exoplanets, but this is potentially risky as our understanding of atmospheric processes under extreme conditions may well be incomplete. 
\item A large sample size of atmospheres may be needed to disentangle the role of a variety of complex physical and chemical effects.  The phase space of exoplanet atmospheres is large.  While stellar atmospheres can largely be understood in terms of the 3 factors of $T_\mathrm{eff}$, surface gravity, and metallicity, a much larger number of factors influences planetary atmospheres \citep{Fortney20}.  While dramatic progress will certainly be made in the 2020s, it will likely open up many new questions that can only be solved from spectra of numerous, possibly hundreds, of exoplanets. 
\item It is becoming increasingly apparent that heterogeneities on the stellar surface can substantially affect transmission spectra, and if not accounted for can result in biased interpretation, \citep[e.g.,][]{Rackham18a,Rackham18b}. Further study and modelling of starspots and faculae, requiring input from the fields of heliophysics and stellar astrophysics, will be required to support transit spectroscopy with \emph{JWST} and beyond. 

\item The challenges for atmospheric retrievals lie in expanding the current paradigm to higher dimensional models and higher quality observations. Robust retrievals involve computing $\sim$ $10^4-10^9$ model evaluations on a $\sim$10-20 dimensional parameter space, depending on the data at hand, even for 1-D models. Therefore, multi-D retrievals are associated with a number of practical challenges. These include increased computation time, requirement of realistic and efficient parameterisations for temperature profiles, compositions and clouds/hazes, and efficient statistical inference and parameter estimation algorithms.

\item New advances are also being made in atmospheric retrievals using a combination of high-resolution and low-resolution spectra. However, there are significant challenges in adapting both modeling and retrieval algorithms for this purpose, including computational costs as well as needs for robust likelihood metrics for the retrievals with multi-instrument, multi-resolution data. 

\item Retrievals of directly-imaged and isolated planets/sub-stellar objects are typically limited by a lack of measured masses and radii, and hence gravities, which need to be retrieved from the spectra in addition to the atmospheric properties. The spectra of such objects are also sensitive to temperature structure in the deep atmosphere as well as clouds, which make their retrievals more challenging than those of thermal emission from irradiated objects. 

\item Since exoplanetary atmospheres probe regimes not seen on Earth or in the solar system, the needed physical and chemical inputs into models can be missing or incomplete.  Needed improvements in areas such as the opacities of molecules, reaction rate constants, and optical properties of aerosols can come from increased investments in laboratory work and first-principle simulations \citep{Fortney19}.

\item Ultimately, interpretation of atmospheric observations will require a holistic approach of using both robust atmospheric retrieval algorithms as well as detailed self-consistent models in tandem to both constrain the atmospheric properties and the underlying physical processes. Thus the ultimate challenge in this area is to consider the wide range of atmospheric processes possible (energy transport, clouds/hazes, disequilibrium chemistry, dynamics, escape) and the broad range of observations (UV to infrared) that can probe the different processes in a unified framework. 

\end{itemize}


\begin{thebibliography}{85}
\expandafter\ifx\csname natexlab\endcsname\relax\def\natexlab#1{#1}\fi

\bibitem[{{Ackerman} \& {Marley}(2001)}]{AM01}
{Ackerman}, A.~S. \& {Marley}, M.~S. 2001, \apj, 556, 872

\bibitem[{Ackerman \& Marley(2001)}]{Ackerman2001}
Ackerman, A.~S. \& Marley, M.~S. 2001, The Astrophysical Journal, 556, 872

\bibitem[{Amundsen {et~al.}(2014)Amundsen, Baraffe, Tremblin, Manners, Hayek,
  Mayne, \& Acreman}]{Amundsen:2014}
Amundsen, D., Baraffe, I., Tremblin, P., Manners, J., Hayek, W., Mayne, N., \&
  Acreman, D. 2014, Astronomy and Astrophysics, 564, A59

\bibitem[{{Amundsen} {et~al.}(2017){Amundsen}, {Tremblin}, {Manners},
  {Baraffe}, \& {Mayne}}]{Amundsen17}
{Amundsen}, D.~S., {Tremblin}, P., {Manners}, J., {Baraffe}, I., \& {Mayne},
  N.~J. 2017, \aap, 598, A97

\bibitem[{{Barman} {et~al.}(2001){Barman}, {Hauschildt}, \&
  {Allard}}]{Barman01}
{Barman}, T.~S., {Hauschildt}, P.~H., \& {Allard}, F. 2001, \apj, 556, 885

\bibitem[{{Barstow}(2020)}]{barstow20b}
{Barstow}, J.~K. 2020, \mnras, 497, 4183

\bibitem[{{Barstow} {et~al.}(2017){Barstow}, {Aigrain}, {Irwin}, \&
  {Sing}}]{Barstow17}
{Barstow}, J.~K., {Aigrain}, S., {Irwin}, P.~G.~J., \& {Sing}, D.~K. 2017,
  \apj, 834, 50

\bibitem[{{Barstow} {et~al.}(2020){Barstow}, {Changeat}, {Garland}, {Line},
  {Rocchetto}, \& {Waldmann}}]{Barstow20}
{Barstow}, J.~K., {Changeat}, Q., {Garland}, R., {Line}, M.~R., {Rocchetto},
  M., \& {Waldmann}, I.~P. 2020, \mnras, 493, 4884

\bibitem[{{Barstow} \& {Irwin}(2016)}]{Barstow16}
{Barstow}, J.~K. \& {Irwin}, P.~G.~J. 2016, \mnras, 461, L92

\bibitem[{{Beichman} {et~al.}(2014){Beichman}, {Benneke}, {Knutson}, {Smith},
  {Lagage}, {Dressing}, {Latham}, {Lunine}, {Birkmann}, {Ferruit}, {Giardino},
  {Kempton}, {Carey}, {Krick}, {Deroo}, {Mandell}, {Ressler}, {Shporer},
  {Swain}, {Vasisht}, {Ricker}, {Bouwman}, {Crossfield}, {Greene}, {Howell},
  {Christiansen}, {Ciardi}, {Clampin}, {Greenhouse}, {Sozzetti}, {Goudfrooij},
  {Hines}, {Keyes}, {Lee}, {McCullough}, {Robberto}, {Stansberry}, {Valenti},
  {Rieke}, {Rieke}, {Fortney}, {Bean}, {Kreidberg}, {Ehrenreich}, {Deming},
  {Albert}, {Doyon}, \& {Sing}}]{Beichman14}
{Beichman}, C., {Benneke}, B., {Knutson}, H., {Smith}, R., {Lagage}, P.-O.,
  {Dressing}, C., {Latham}, D., {Lunine}, J., {Birkmann}, S., {Ferruit}, P.,
  {Giardino}, G., {Kempton}, E., {Carey}, S., {Krick}, J., {Deroo}, P.~D.,
  {Mandell}, A., {Ressler}, M.~E., {Shporer}, A., {Swain}, M., {Vasisht}, G.,
  {Ricker}, G., {Bouwman}, J., {Crossfield}, I., {Greene}, T., {Howell}, S.,
  {Christiansen}, J., {Ciardi}, D., {Clampin}, M., {Greenhouse}, M.,
  {Sozzetti}, A., {Goudfrooij}, P., {Hines}, D., {Keyes}, T., {Lee}, J.,
  {McCullough}, P., {Robberto}, M., {Stansberry}, J., {Valenti}, J., {Rieke},
  M., {Rieke}, G., {Fortney}, J., {Bean}, J., {Kreidberg}, L., {Ehrenreich},
  D., {Deming}, D., {Albert}, L., {Doyon}, R., \& {Sing}, D. 2014, \pasp, 126,
  1134

\bibitem[{{Birkby}(2018)}]{Birkby18}
{Birkby}, J.~L. 2018, arXiv e-prints, arXiv:1806.04617

\bibitem[{{Blecic} {et~al.}(2017){Blecic}, {Dobbs-Dixon}, \&
  {Greene}}]{Blecic17}
{Blecic}, J., {Dobbs-Dixon}, I., \& {Greene}, T. 2017, \apj, 848, 127

\bibitem[{{Brandt} {et~al.}(2019){Brandt}, {Briesemeister}, {Savransky},
  {Fitzgerald}, {Mazin}, {Fortney}, {Dupuy}, {Bowler}, {Sallum}, {Mawet},
  {Skemer}, {Vasisht}, {Miller-Blanchard}, {Wang}, {Guyon}, {Meshkat},
  {Jensen-Clem}, {Serabyn}, {Ruane}, {Liu}, {Jovanovic}, {Morley}, {Perrin},
  {McElwain}, {Roberge}, {Girard}, {Close}, {Ngo}, {Marley}, {Bendek}, {Ragland
  }, \& {Pueyo}}]{Brandt19}
{Brandt}, T., {Briesemeister}, Z., {Savransky}, D., {Fitzgerald}, M., {Mazin},
  B., {Fortney}, J., {Dupuy}, T., {Bowler}, B., {Sallum}, S., {Mawet}, D.,
  {Skemer}, A., {Vasisht}, G., {Miller-Blanchard}, M., {Wang}, J., {Guyon}, O.,
  {Meshkat}, T., {Jensen-Clem}, R., {Serabyn}, E., {Ruane}, G., {Liu}, M.,
  {Jovanovic}, N., {Morley}, C., {Perrin}, M., {McElwain}, M., {Roberge}, A.,
  {Girard}, J., {Close}, L., {Ngo}, H., {Marley}, M., {Bendek}, E., {Ragland },
  S., \& {Pueyo}, L. 2019, \baas, 51, 269

\bibitem[{Brogi {et~al.}(2016)Brogi, de~Kok, Albrecht, Snellen, Birkby, \&
  Schwarz}]{Brogi:2015}
Brogi, M., de~Kok, R., Albrecht, S., Snellen, I., Birkby, J., \& Schwarz, H.
  2016, The Astrophysical Journal, 817, 106

\bibitem[{{Burrows} {et~al.}(1997){Burrows}, {Marley}, {Hubbard}, {Lunine},
  {Guillot}, {Saumon}, {Freedman}, {Sudarsky}, \& {Sharp}}]{Burrows97}
{Burrows}, A., {Marley}, M., {Hubbard}, W.~B., {Lunine}, J.~I., {Guillot}, T.,
  {Saumon}, D., {Freedman}, R., {Sudarsky}, D., \& {Sharp}, C. 1997, \apj, 491,
  856

\bibitem[{{Caldas} {et~al.}(2019){Caldas}, {Leconte}, {Selsis}, {Waldmann},
  {Bord{\'e}}, {Rocchetto}, \& {Charnay}}]{Caldas19}
{Caldas}, A., {Leconte}, J., {Selsis}, F., {Waldmann}, I.~P., {Bord{\'e}}, P.,
  {Rocchetto}, M., \& {Charnay}, B. 2019, \aap, 623, A161

\bibitem[{{Carone} {et~al.}(2014){Carone}, {Keppens}, \& {Decin}}]{Carone14}
{Carone}, L., {Keppens}, R., \& {Decin}, L. 2014, \mnras, 445, 930

\bibitem[{{Changeat} {et~al.}(2020){Changeat}, {Al-Refaie}, {Mugnai},
  {Edwards}, {Waldmann}, {Pascale}, \& {Tinetti}}]{changeat20}
{Changeat}, Q., {Al-Refaie}, A., {Mugnai}, L.~V., {Edwards}, B., {Waldmann},
  I.~P., {Pascale}, E., \& {Tinetti}, G. 2020, \aj, 160, 80

\bibitem[{{Deitrick} {et~al.}(2020){Deitrick}, {Mendon{\c{c}}a},
  {Schroffenegger}, {Grimm}, {Tsai}, \& {Heng}}]{Deitrick20}
{Deitrick}, R., {Mendon{\c{c}}a}, J.~M., {Schroffenegger}, U., {Grimm}, S.~L.,
  {Tsai}, S.-M., \& {Heng}, K. 2020, \apjs, 248, 30

\bibitem[{Drummond {et~al.}(2018)Drummond, Mayne, Manners, \& {et
  al.}}]{Drummond:2018aa}
Drummond, B., Mayne, N., Manners, J., \& {et al.} 2018, The Astrophysical
  Journal, 869, 28

\bibitem[{{Feng} {et~al.}(2020){Feng}, {Line}, \& {Fortney}}]{Feng20}
{Feng}, Y.~K., {Line}, M.~R., \& {Fortney}, J.~J. 2020, \aj, 160, 137

\bibitem[{{Feng} {et~al.}(2016){Feng}, {Line}, {Fortney}, {Stevenson}, {Bean},
  {Kreidberg}, \& {Parmentier}}]{Feng16}
{Feng}, Y.~K., {Line}, M.~R., {Fortney}, J.~J., {Stevenson}, K.~B., {Bean}, J.,
  {Kreidberg}, L., \& {Parmentier}, V. 2016, \apj, 829, 52

\bibitem[{{Fortney} {et~al.}(2019){Fortney}, {Robinson}, {Domagal-Goldman},
  {Genio}, {Gordon}, {Gharib-Nezhad}, {Lewis}, {Sousa-Silva}, {Airapetian},
  {Drouin}, {Hargreaves}, {Huang}, {Karman}, {Ramirez}, {Rieker}, {Tennyson},
  {Wordsworth}, {Yurchenko}, {Johnson}, {Lee}, {Marley}, {Dong}, {Kane},
  {L{\'o}pez-Morales}, {Fauchez}, {Lee}, {Sung}, {Haghighipour}, {Horst},
  {Gao}, {Kao}, {Dressing}, {Lupu}, {Savin}, {Fleury}, {Venot}, {Ascenzi},
  {Milam}, {Linnartz}, {Gudipati}, {Gronoff}, {Salama}, {Gavilan}, {Bouwman},
  {Turbet}, {Benilan}, {Henderson}, {Batalha}, {Jensen-Clem}, {Lyons},
  {Freedman}, {Schwieterman}, {Goyal}, {Mancini}, {Irwin}, {Desert},
  {Molaverdikhani}, {Gizis}, {Taylor}, {Lothringer}, {Pierrehumbert}, {Zellem},
  {Batalha}, {Rugheimer}, {Lustig-Yaeger}, {Hu}, {Kempton}, {Arney}, {Line},
  {Alam}, {Moses}, {Iro}, {Kreidberg}, {Blecic}, {Louden}, {Molli{\`e}re},
  {Stevenson}, {Swain}, {Bott}, {Madhusudhan}, {Krissansen-Totton}, {Deming},
  {Kitiashvili}, {Shkolnik}, {Rustamkulov}, {Rogers}, \& {Close}}]{Fortney19}
{Fortney}, J., {Robinson}, T.~D., {Domagal-Goldman}, S., {Genio}, A. D.~D.,
  {Gordon}, I.~E., {Gharib-Nezhad}, E., {Lewis}, N., {Sousa-Silva}, C.,
  {Airapetian}, V., {Drouin}, B., {Hargreaves}, R.~J., {Huang}, X., {Karman},
  T., {Ramirez}, R.~M., {Rieker}, G.~B., {Tennyson}, J., {Wordsworth}, R.,
  {Yurchenko}, S.~N., {Johnson}, A. r.~V., {Lee}, T.~J., {Marley}, M.~S.,
  {Dong}, C., {Kane}, S., {L{\'o}pez-Morales}, M., {Fauchez}, T., {Lee}, T.,
  {Sung}, K., {Haghighipour}, N., {Horst}, S., {Gao}, P., {Kao}, D.-y.,
  {Dressing}, C., {Lupu}, R., {Savin}, D.~W., {Fleury}, B., {Venot}, O.,
  {Ascenzi}, D., {Milam}, S., {Linnartz}, H., {Gudipati}, M., {Gronoff}, G.,
  {Salama}, F., {Gavilan}, L., {Bouwman}, J., {Turbet}, M., {Benilan}, Y.,
  {Henderson}, B., {Batalha}, N., {Jensen-Clem}, R., {Lyons}, T., {Freedman},
  R., {Schwieterman}, E., {Goyal}, J., {Mancini}, L., {Irwin}, P., {Desert},
  J.-M., {Molaverdikhani}, K., {Gizis}, J., {Taylor}, J., {Lothringer}, J.,
  {Pierrehumbert}, R., {Zellem}, R., {Batalha}, N., {Rugheimer}, S.,
  {Lustig-Yaeger}, J., {Hu}, R., {Kempton}, E., {Arney}, G., {Line}, M.,
  {Alam}, M., {Moses}, J., {Iro}, N., {Kreidberg}, L., {Blecic}, J., {Louden},
  T., {Molli{\`e}re}, P., {Stevenson}, K., {Swain}, M., {Bott}, K.,
  {Madhusudhan}, N., {Krissansen-Totton}, J., {Deming}, D., {Kitiashvili}, I.,
  {Shkolnik}, E., {Rustamkulov}, Z., {Rogers}, L., \& {Close}, L. 2019,
  Astro2020: Decadal Survey on Astronomy and Astrophysics, 2020, 146

\bibitem[{{Fortney} {et~al.}(2008){Fortney}, {Lodders}, {Marley}, \&
  {Freedman}}]{Fortney08a}
{Fortney}, J.~J., {Lodders}, K., {Marley}, M.~S., \& {Freedman}, R.~S. 2008,
  \apj, 678, 1419

\bibitem[{{Fortney} {et~al.}(2020){Fortney}, {Visscher}, {Marley}, {Hood},
  {Line}, {Thorngren}, {Freedman}, \& {Lupu}}]{Fortney20}
{Fortney}, J.~J., {Visscher}, C., {Marley}, M.~S., {Hood}, C.~E., {Line},
  M.~R., {Thorngren}, D.~P., {Freedman}, R.~S., \& {Lupu}, R. 2020, \aj, 160,
  288

\bibitem[{{Gandhi} \& {Madhusudhan}(2017)}]{Gandhi17}
{Gandhi}, S. \& {Madhusudhan}, N. 2017, \mnras, 472, 2334

\bibitem[{Gao {et~al.}(2020)Gao, Thorngren, Lee, \& {et al.}}]{Gao:2020aa}
Gao, P., Thorngren, D., Lee, G., \& {et al.} 2020, Nature Astronomy

\bibitem[{{Goody} {et~al.}(1989){Goody}, {West}, {Chen}, \& {Crisp}}]{Goody89}
{Goody}, R., {West}, R., {Chen}, L., \& {Crisp}, D. 1989, Journal of
  Quantitative Spectroscopy and Radiative Transfer, 42, 539

\bibitem[{{Goukenleuque} {et~al.}(2000){Goukenleuque}, {B{\'e}zard}, {Joguet},
  {Lellouch}, \& {Freedman}}]{Gouk00}
{Goukenleuque}, C., {B{\'e}zard}, B., {Joguet}, B., {Lellouch}, E., \&
  {Freedman}, R. 2000, Icarus, 143, 308

\bibitem[{{Gravity Collaboration} {et~al.}(2020){Gravity Collaboration},
  {Nowak}, {Lacour}, {Molli{\`e}re}, {Wang}, {Charnay}, {van Dishoeck},
  {Abuter}, {Amorim}, {Berger}, {Beust}, {Bonnefoy}, {Bonnet}, {Brandner},
  {Buron}, {Cantalloube}, {Collin}, {Chapron}, {Cl{\'e}net}, {Coud{\'e} Du
  Foresto}, {de Zeeuw}, {Dembet}, {Dexter}, {Duvert}, {Eckart}, {Eisenhauer},
  {F{\"o}rster Schreiber}, {F{\'e}dou}, {Garcia Lopez}, {Gao}, {Gendron},
  {Genzel}, {Gillessen}, {Hau{\ss}mann}, {Henning}, {Hippler}, {Hubert},
  {Jocou}, {Kervella}, {Lagrange}, {Lapeyr{\`e}re}, {Le Bouquin}, {L{\'e}na},
  {Maire}, {Ott}, {Paumard}, {Paladini}, {Perraut}, {Perrin}, {Pueyo}, {Pfuhl},
  {Rabien}, {Rau}, {Rodr{\'\i}guez-Coira}, {Rousset}, {Scheithauer},
  {Shangguan}, {Straub}, {Straubmeier}, {Sturm}, {Tacconi}, {Vincent},
  {Widmann}, {Wieprecht}, {Wiezorrek}, {Woillez}, {Yazici}, \&
  {Ziegler}}]{Gravity20}
{Gravity Collaboration}, {Nowak}, M., {Lacour}, S., {Molli{\`e}re}, P., {Wang},
  J., {Charnay}, B., {van Dishoeck}, E.~F., {Abuter}, R., {Amorim}, A.,
  {Berger}, J.~P., {Beust}, H., {Bonnefoy}, M., {Bonnet}, H., {Brandner}, W.,
  {Buron}, A., {Cantalloube}, F., {Collin}, C., {Chapron}, F., {Cl{\'e}net},
  Y., {Coud{\'e} Du Foresto}, V., {de Zeeuw}, P.~T., {Dembet}, R., {Dexter},
  J., {Duvert}, G., {Eckart}, A., {Eisenhauer}, F., {F{\"o}rster Schreiber},
  N.~M., {F{\'e}dou}, P., {Garcia Lopez}, R., {Gao}, F., {Gendron}, E.,
  {Genzel}, R., {Gillessen}, S., {Hau{\ss}mann}, F., {Henning}, T., {Hippler},
  S., {Hubert}, Z., {Jocou}, L., {Kervella}, P., {Lagrange}, A.~M.,
  {Lapeyr{\`e}re}, V., {Le Bouquin}, J.~B., {L{\'e}na}, P., {Maire}, A.~L.,
  {Ott}, T., {Paumard}, T., {Paladini}, C., {Perraut}, K., {Perrin}, G.,
  {Pueyo}, L., {Pfuhl}, O., {Rabien}, S., {Rau}, C., {Rodr{\'\i}guez-Coira},
  G., {Rousset}, G., {Scheithauer}, S., {Shangguan}, J., {Straub}, O.,
  {Straubmeier}, C., {Sturm}, E., {Tacconi}, L.~J., {Vincent}, F., {Widmann},
  F., {Wieprecht}, E., {Wiezorrek}, E., {Woillez}, J., {Yazici}, S., \&
  {Ziegler}, D. 2020, \aap, 633, A110

\bibitem[{{Greene} {et~al.}(2016){Greene}, {Line}, {Montero}, {Fortney},
  {Lustig-Yaeger}, \& {Luther}}]{Greene16}
{Greene}, T.~P., {Line}, M.~R., {Montero}, C., {Fortney}, J.~J.,
  {Lustig-Yaeger}, J., \& {Luther}, K. 2016, \apj, 817, 17

\bibitem[{{He} {et~al.}(2020){He}, {H{\"o}rst}, {Lewis}, {Yu}, {Moses},
  {McGuiggan}, {Marley}, {Kempton}, {Moran}, {Morley}, \& {Vuitton}}]{Chao20}
{He}, C., {H{\"o}rst}, S.~M., {Lewis}, N.~K., {Yu}, X., {Moses}, J.~I.,
  {McGuiggan}, P., {Marley}, M.~S., {Kempton}, E. M.~R., {Moran}, S.~E.,
  {Morley}, C.~V., \& {Vuitton}, V. 2020, Nature Astronomy

\bibitem[{{Helling} \& {Casewell}(2014)}]{Helling14}
{Helling}, C. \& {Casewell}, S. 2014, \aapr, 22, 80

\bibitem[{Helling {et~al.}(2016)Helling, Lee, Mayne, Amundsen, Khaimova, Unger,
  Manners, Acreman, \& Smith}]{Helling:2016}
Helling, C., Lee, G.; Dobbs-Dixon, I., Mayne, N., Amundsen, D.~S., Khaimova,
  J., Unger, A.~A., Manners, J., Acreman, D., \& Smith, C. 2016, Monthly
  Notices of the Royal Astronomical Society, 460, 855

\bibitem[{{Helling} {et~al.}(2017){Helling}, {Tootill}, {Woitke}, \&
  {Lee}}]{Helling17}
{Helling}, C., {Tootill}, D., {Woitke}, P., \& {Lee}, G. 2017, \aap, 603, A123

\bibitem[{{Heng} \& {Showman}(2015)}]{Heng15}
{Heng}, K. \& {Showman}, A.~P. 2015, Annual Review of Earth and Planetary
  Sciences, 43, 509

\bibitem[{{H{\"o}rst} {et~al.}(2018){H{\"o}rst}, {He}, {Lewis}, {Kempton},
  {Marley}, {Morley}, {Moses}, {Valenti}, \& {Vuitton}}]{Horst18}
{H{\"o}rst}, S.~M., {He}, C., {Lewis}, N.~K., {Kempton}, E. M.~R., {Marley},
  M.~S., {Morley}, C.~V., {Moses}, J.~I., {Valenti}, J.~A., \& {Vuitton}, V.
  2018, Nature Astronomy, 2, 303

\bibitem[{{Irwin} {et~al.}(2020){Irwin}, {Parmentier}, {Taylor}, {Barstow},
  {Aigrain}, {Lee}, \& {Garland }}]{Irwin20}
{Irwin}, P. G.~J., {Parmentier}, V., {Taylor}, J., {Barstow}, J., {Aigrain},
  S., {Lee}, G. K.~H., \& {Garland }, R. 2020, \mnras, 493, 106

\bibitem[{{Irwin} {et~al.}(2008){Irwin}, {Teanby}, {de Kok}, {Fletcher},
  {Howett}, {Tsang}, {Wilson}, {Calcutt}, {Nixon}, \& {Parrish}}]{Irwin08}
{Irwin}, P.~G.~J., {Teanby}, N.~A., {de Kok}, R., {Fletcher}, L.~N., {Howett},
  C.~J.~A., {Tsang}, C.~C.~C., {Wilson}, C.~F., {Calcutt}, S.~B., {Nixon},
  C.~A., \& {Parrish}, P.~D. 2008, \jqsrt, 109, 1136

\bibitem[{{Joshi} {et~al.}(1997){Joshi}, {Haberle}, \& {Reynolds}}]{Joshi97}
{Joshi}, M.~M., {Haberle}, R.~M., \& {Reynolds}, R.~T. 1997, \icarus, 129, 450

\bibitem[{{Kite} {et~al.}(2020){Kite}, {Fegley}, {Schaefer}, \&
  {Ford}}]{Kite20}
{Kite}, E.~S., {Fegley}, Bruce, J., {Schaefer}, L., \& {Ford}, E.~B. 2020,
  \apj, 891, 111

\bibitem[{{Kitzmann} {et~al.}(2020){Kitzmann}, {Heng}, {Oreshenko}, {Grimm},
  {Apai}, {Bowler}, {Burgasser}, \& {Marley}}]{kitzmann20}
{Kitzmann}, D., {Heng}, K., {Oreshenko}, M., {Grimm}, S.~L., {Apai}, D.,
  {Bowler}, B.~P., {Burgasser}, A.~J., \& {Marley}, M.~S. 2020, \apj, 890, 174

\bibitem[{{Koll} {et~al.}(2019){Koll}, {Malik}, {Mansfield}, {Kempton}, {Kite},
  {Abbot}, \& {Bean}}]{Koll19}
{Koll}, D. D.~B., {Malik}, M., {Mansfield}, M., {Kempton}, E. M.~R., {Kite},
  E., {Abbot}, D., \& {Bean}, J.~L. 2019, \apj, 886, 140

\bibitem[{{Kreidberg} {et~al.}(2018){Kreidberg}, {Line}, {Thorngren}, {Morley},
  \& {Stevenson}}]{Kreidberg18}
{Kreidberg}, L., {Line}, M.~R., {Thorngren}, D., {Morley}, C.~V., \&
  {Stevenson}, K.~B. 2018, \apjl, 858, L6

\bibitem[{{Lacy} \& {Burrows}(2020)}]{Lacy20}
{Lacy}, B.~I. \& {Burrows}, A. 2020, \apj, 905, 131

\bibitem[{{Leconte} \& {Chabrier}(2013)}]{Leconte13}
{Leconte}, J. \& {Chabrier}, G. 2013, Nature Geoscience, 6, 347

\bibitem[{Lee {et~al.}(2016)Lee, Dobbs-Dixon, Helling, Bognar, \&
  Woitke}]{Lee:2016}
Lee, G., Dobbs-Dixon, I., Helling, C., Bognar, K., \& Woitke, P. 2016,
  Astronomy and Astrophysics, 594, A48

\bibitem[{{Line} {et~al.}(2017){Line}, {Marley}, {Liu}, {Burningham}, {Morley},
  {Hinkel}, {Teske}, {Fortney}, {Freedman}, \& {Lupu}}]{Line17}
{Line}, M.~R., {Marley}, M.~S., {Liu}, M.~C., {Burningham}, B., {Morley},
  C.~V., {Hinkel}, N.~R., {Teske}, J., {Fortney}, J.~J., {Freedman}, R., \&
  {Lupu}, R. 2017, \apj, 848, 83

\bibitem[{{Line} \& {Parmentier}(2016)}]{Line16}
{Line}, M.~R. \& {Parmentier}, V. 2016, \apj, 820, 78

\bibitem[{Line \& Parmentier(2016)}]{Line2016}
Line, M.~R. \& Parmentier, V. 2016, The Astrophysical Journal, 820, 78

\bibitem[{{Line} {et~al.}(2013){Line}, {Wolf}, {Zhang}, {Knutson}, {Kammer},
  {Ellison}, {Deroo}, {Crisp}, \& {Yung}}]{Line13}
{Line}, M.~R., {Wolf}, A.~S., {Zhang}, X., {Knutson}, H., {Kammer}, J.~A.,
  {Ellison}, E., {Deroo}, P., {Crisp}, D., \& {Yung}, Y.~L. 2013, \apj, 775,
  137

\bibitem[{Lines {et~al.}(2018)Lines, Mayne, Boutle, Manners, Lee, Helling,
  Drummond, Amundsen, Goyal, Acreman, Tremblin, \& Kerslake}]{Lines:2018}
Lines, S., Mayne, N., Boutle, I., Manners, J., Lee, G., Helling, C., Drummond,
  B., Amundsen, D., Goyal, J., Acreman, D., Tremblin, P., \& Kerslake, M. 2018,
  Astronomy {\&} Astrophysics, 615, A97

\bibitem[{{Lines} {et~al.}(2019){Lines}, {Mayne}, {Manners}, {Boutle},
  {Drummond}, {Mikal-Evans}, {Kohary}, \& {Sing}}]{lines:2019}
{Lines}, S., {Mayne}, N.~J., {Manners}, J., {Boutle}, I.~A., {Drummond}, B.,
  {Mikal-Evans}, T., {Kohary}, K., \& {Sing}, D.~K. 2019, \mnras, 488, 1332

\bibitem[{{Lodders} \& {Fegley}(2002)}]{Lodders02}
{Lodders}, K. \& {Fegley}, B. 2002, Icarus, 155, 393

\bibitem[{Louden \& Wheatley(2015)}]{Louden:2015}
Louden, T. \& Wheatley, P. 2015, The Astrophysical Journal Letters, 814, L24

\bibitem[{{MacDonald} {et~al.}(2020){MacDonald}, {Goyal}, \&
  {Lewis}}]{MacDonald20}
{MacDonald}, R.~J., {Goyal}, J.~M., \& {Lewis}, N.~K. 2020, \apjl, 893, L43

\bibitem[{{Madhusudhan}(2018)}]{Mad18x}
{Madhusudhan}, N. 2018, arXiv e-prints, arXiv:1808.04824

\bibitem[{{Madhusudhan} {et~al.}(2014){Madhusudhan}, {Amin}, \&
  {Kennedy}}]{Madhu14}
{Madhusudhan}, N., {Amin}, M.~A., \& {Kennedy}, G.~M. 2014, \apjl, 794, L12

\bibitem[{{Madhusudhan} {et~al.}(2011){Madhusudhan}, {Harrington}, {Stevenson},
  {Nymeyer}, {Campo}, {Wheatley}, {Deming}, {Blecic}, {Hardy}, {Lust},
  {Anderson}, {Collier-Cameron}, {Britt}, {Bowman}, {Hebb}, {Hellier},
  {Maxted}, {Pollacco}, \& {West}}]{Madhu11}
{Madhusudhan}, N., {Harrington}, J., {Stevenson}, K.~B., {Nymeyer}, S.,
  {Campo}, C.~J., {Wheatley}, P.~J., {Deming}, D., {Blecic}, J., {Hardy},
  R.~A., {Lust}, N.~B., {Anderson}, D.~R., {Collier-Cameron}, A., {Britt},
  C.~B.~T., {Bowman}, W.~C., {Hebb}, L., {Hellier}, C., {Maxted}, P.~F.~L.,
  {Pollacco}, D., \& {West}, R.~G. 2011, \nat, 469, 64

\bibitem[{Madhusudhan \& Seager(2009)}]{Madhusudhan:2009}
Madhusudhan, N. \& Seager, S. 2009, The Astrophysical Journal, 707, 24

\bibitem[{{Mai} \& {Line}(2019)}]{mai19}
{Mai}, C. \& {Line}, M.~R. 2019, \apj, 883, 144

\bibitem[{{Mayne} {et~al.}(2014){Mayne}, {Baraffe}, {Acreman}, {Smith},
  {Browning}, {Sk{\aa}lid Amundsen}, {Wood}, {Thuburn}, \& {Jackson}}]{Mayne14}
{Mayne}, N.~J., {Baraffe}, I., {Acreman}, D.~M., {Smith}, C., {Browning},
  M.~K., {Sk{\aa}lid Amundsen}, D., {Wood}, N., {Thuburn}, J., \& {Jackson},
  D.~R. 2014, \aap, 561, A1

\bibitem[{{Molli{\`e}re} {et~al.}(2015){Molli{\`e}re}, {van Boekel},
  {Dullemond}, {Henning}, \& {Mordasini}}]{Molliere15}
{Molli{\`e}re}, P., {van Boekel}, R., {Dullemond}, C., {Henning}, T., \&
  {Mordasini}, C. 2015, \apj, 813, 47

\bibitem[{{Mordasini} {et~al.}(2016){Mordasini}, {van Boekel}, {Molli{\`e}re},
  {Henning}, \& {Benneke}}]{Mordasini16}
{Mordasini}, C., {van Boekel}, R., {Molli{\`e}re}, P., {Henning}, T., \&
  {Benneke}, B. 2016, \apj, 832, 41

\bibitem[{{Morley} {et~al.}(2017){Morley}, {Kreidberg}, {Rustamkulov},
  {Robinson}, \& {Fortney}}]{Morley17b}
{Morley}, C.~V., {Kreidberg}, L., {Rustamkulov}, Z., {Robinson}, T., \&
  {Fortney}, J.~J. 2017, \apj, 850, 121

\bibitem[{{Moses} {et~al.}(2011){Moses}, {Visscher}, {Fortney}, {Showman},
  {Lewis}, {Griffith}, {Klippenstein}, {Shabram}, {Friedson}, {Marley}, \&
  {Freedman}}]{Moses11}
{Moses}, J.~I., {Visscher}, C., {Fortney}, J.~J., {Showman}, A.~P., {Lewis},
  N.~K., {Griffith}, C.~A., {Klippenstein}, S.~J., {Shabram}, M., {Friedson},
  A.~J., {Marley}, M.~S., \& {Freedman}, R.~S. 2011, \apj, 737, 15

\bibitem[{{{\"O}berg} {et~al.}(2011){{\"O}berg}, {Murray-Clay}, \&
  {Bergin}}]{Oberg11}
{{\"O}berg}, K.~I., {Murray-Clay}, R., \& {Bergin}, E.~A. 2011, \apjl, 743, L16

\bibitem[{{Rackham} {et~al.}(2018){Rackham}, {Apai}, \&
  {Giampapa}}]{Rackham18a}
{Rackham}, B.~V., {Apai}, D., \& {Giampapa}, M.~S. 2018, \apj, 853, 122

\bibitem[{{Rackham} {et~al.}(2019){Rackham}, {Apai}, \&
  {Giampapa}}]{Rackham18b}
---. 2019, \aj, 157, 96

\bibitem[{Rauscher \& Menou(2010)}]{Rauscher:2010}
Rauscher, E. \& Menou, K. 2010, The Astrophysical Journal, 714, 1334

\bibitem[{{Rodgers}(2000)}]{Rodgers00}
{Rodgers}, C.~D. 2000, {Inverse Methods for Atmospheric Sounding: Theory and
  Practice}

\bibitem[{Roman \& Rauscher(2019)}]{Roman:2019aa}
Roman, M. \& Rauscher, E. 2019, The Astrophysical Journal, 872, 1

\bibitem[{{Seager} \& {Sasselov}(1998)}]{SS98}
{Seager}, S. \& {Sasselov}, D.~D. 1998, \apjl, 502, L157

\bibitem[{{Showman} {et~al.}(2009){Showman}, {Fortney}, {Lian}, {Marley},
  {Freedman}, {Knutson}, \& {Charbonneau}}]{Showman09}
{Showman}, A.~P., {Fortney}, J.~J., {Lian}, Y., {Marley}, M.~S., {Freedman},
  R.~S., {Knutson}, H.~A., \& {Charbonneau}, D. 2009, \apj, 699, 564

\bibitem[{{Showman} \& {Guillot}(2002)}]{Showman02}
{Showman}, A.~P. \& {Guillot}, T. 2002, \aap, 385, 166

\bibitem[{Snellen {et~al.}(2010)Snellen, de~Kok, de~Mooij, \&
  Albrecht}]{Snellen:2010}
Snellen, I., de~Kok, R., de~Mooij, E., \& Albrecht, S. 2010, Nature, 465, 1049

\bibitem[{Stevenson {et~al.}(2014)Stevenson, Desert, Line, Bean, Fortney,
  Showman, Kataria, Kriedberg, McCullough, Henry, Charbonneau, Burrows, Seager,
  Madhusudhan, Williamson, \& Homeier}]{Stevenson:2014}
Stevenson, K., Desert, J., Line, M., Bean, J., Fortney, J., Showman, A.,
  Kataria, T., Kriedberg, L., McCullough, P., Henry, G., Charbonneau, D.,
  Burrows, A., Seager, S., Madhusudhan, N., Williamson, M., \& Homeier, D.
  2014, Science, 346, 838

\bibitem[{{Taylor} {et~al.}(2020){Taylor}, {Parmentier}, {Irwin}, {Aigrain},
  {Lee}, \& {Krissansen-Totton}}]{Taylor20}
{Taylor}, J., {Parmentier}, V., {Irwin}, P. G.~J., {Aigrain}, S., {Lee}, G.
  K.~H., \& {Krissansen-Totton}, J. 2020, \mnras, 493, 4342

\bibitem[{{Tinetti} {et~al.}(2018){Tinetti}, {Drossart}, {Eccleston},
  {Hartogh}, {Heske}, {Leconte}, {Micela}, {Ollivier}, {Pilbratt}, {Puig},
  {Turrini}, {Vandenbussche}, {Wolkenberg}, {Beaulieu}, {Buchave}, {Ferus},
  {Griffin}, {Guedel}, {Justtanont}, {Lagage}, {Machado}, {Malaguti}, {Min},
  {N{\o}rgaard-Nielsen}, {Rataj}, {Ray}, {Ribas}, {Swain}, {Szabo}, {Werner},
  {Barstow}, {Burleigh}, {Cho}, {du Foresto}, {Coustenis}, {Decin}, {Encrenaz},
  {Galand }, {Gillon}, {Helled}, {Morales}, {Mu{\~n}oz}, {Moneti}, {Pagano},
  {Pascale}, {Piccioni}, {Pinfield}, {Sarkar}, {Selsis}, {Tennyson}, {Triaud},
  {Venot}, {Waldmann}, {Waltham}, {Wright}, {Amiaux}, {Augu{\`e}res},
  {Berth{\'e}}, {Bezawada}, {Bishop}, {Bowles}, {Coffey}, {Colom{\'e}},
  {Crook}, {Crouzet}, {Da Peppo}, {Sanz}, {Focardi}, {Frericks}, {Hunt},
  {Kohley}, {Middleton}, {Morgante}, {Ottensamer}, {Pace}, {Pearson},
  {Stamper}, {Symonds}, {Rengel}, {Renotte}, {Ade}, {Affer}, {Alard}, {Allard},
  {Altieri}, {Andr{\'e}}, {Arena}, {Argyriou}, {Aylward}, {Baccani}, {Bakos},
  {Banaszkiewicz}, {Barlow}, {Batista}, {Bellucci}, {Benatti}, {Bernardi},
  {B{\'e}zard}, {Blecka}, {Bolmont}, {Bonfond}, {Bonito}, {Bonomo}, {Brucato},
  {Brun}, {Bryson}, {Bujwan}, {Casewell}, {Charnay}, {Pestellini}, {Chen},
  {Ciaravella}, {Claudi}, {Cl{\'e}dassou}, {Damasso}, {Damiano}, {Danielski},
  {Deroo}, {Di Giorgio}, {Dominik}, {Doublier}, {Doyle}, {Doyon}, {Drummond},
  {Duong}, {Eales}, {Edwards}, {Farina}, {Flaccomio}, {Fletcher}, {Forget},
  {Fossey}, {Fr{\"a}nz}, {Fujii}, {Garc{\'\i}a-Piquer}, {Gear}, {Geoffray},
  {G{\'e}rard}, {Gesa}, {Gomez}, {Graczyk}, {Griffith}, {Grodent}, {Guarcello},
  {Gustin}, {Hamano}, {Hargrave}, {Hello}, {Heng}, {Herrero}, {Hornstrup},
  {Hubert}, {Ida}, {Ikoma}, {Iro}, {Irwin}, {Jarchow}, {Jaubert}, {Jones},
  {Julien}, {Kameda}, {Kerschbaum}, {Kervella}, {Koskinen}, {Krijger}, {Krupp},
  {Lafarga}, {Landini}, {Lellouch}, {Leto}, {Luntzer}, {Rank-L{\"u}ftinger},
  {Maggio}, {Maldonado}, {Maillard}, {Mall}, {Marquette}, {Mathis}, {Maxted},
  {Matsuo}, {Medvedev}, {Miguel}, {Minier}, {Morello}, {Mura}, {Narita},
  {Nascimbeni}, {Nguyen Tong}, {Noce}, {Oliva}, {Palle}, {Palmer}, {Pancrazzi},
  {Papageorgiou}, {Parmentier}, {Perger}, {Petralia}, {Pezzuto},
  {Pierrehumbert}, {Pillitteri}, {Piotto}, {Pisano}, {Prisinzano}, {Radioti},
  {R{\'e}ess}, {Rezac}, {Rocchetto}, {Rosich}, {Sanna}, {Santerne}, {Savini},
  {Scandariato}, {Sicardy}, {Sierra}, {Sindoni}, {Skup}, {Snellen}, {Sobiecki},
  {Soret}, {Sozzetti}, {Stiepen}, {Strugarek}, {Taylor}, {Taylor}, {Terenzi},
  {Tessenyi}, {Tsiaras}, {Tucker}, {Valencia}, {Vasisht}, {Vazan}, {Vilardell},
  {Vinatier}, {Viti}, {Waters}, {Wawer}, {Wawrzaszek}, {Whitworth}, {Yung},
  {Yurchenko}, {Osorio}, {Zellem}, {Zingales}, \& {Zwart}}]{Tinetti18}
{Tinetti}, G., {Drossart}, P., {Eccleston}, P., {Hartogh}, P., {Heske}, A.,
  {Leconte}, J., {Micela}, G., {Ollivier}, M., {Pilbratt}, G., {Puig}, L.,
  {Turrini}, D., {Vandenbussche}, B., {Wolkenberg}, P., {Beaulieu}, J.-P.,
  {Buchave}, L.~A., {Ferus}, M., {Griffin}, M., {Guedel}, M., {Justtanont}, K.,
  {Lagage}, P.-O., {Machado}, P., {Malaguti}, G., {Min}, M.,
  {N{\o}rgaard-Nielsen}, H.~U., {Rataj}, M., {Ray}, T., {Ribas}, I., {Swain},
  M., {Szabo}, R., {Werner}, S., {Barstow}, J., {Burleigh}, M., {Cho}, J., {du
  Foresto}, V.~C., {Coustenis}, A., {Decin}, L., {Encrenaz}, T., {Galand }, M.,
  {Gillon}, M., {Helled}, R., {Morales}, J.~C., {Mu{\~n}oz}, A.~G., {Moneti},
  A., {Pagano}, I., {Pascale}, E., {Piccioni}, G., {Pinfield}, D., {Sarkar},
  S., {Selsis}, F., {Tennyson}, J., {Triaud}, A., {Venot}, O., {Waldmann}, I.,
  {Waltham}, D., {Wright}, G., {Amiaux}, J., {Augu{\`e}res}, J.-L.,
  {Berth{\'e}}, M., {Bezawada}, N., {Bishop}, G., {Bowles}, N., {Coffey}, D.,
  {Colom{\'e}}, J., {Crook}, M., {Crouzet}, P.-E., {Da Peppo}, V., {Sanz},
  I.~E., {Focardi}, M., {Frericks}, M., {Hunt}, T., {Kohley}, R., {Middleton},
  K., {Morgante}, G., {Ottensamer}, R., {Pace}, E., {Pearson}, C., {Stamper},
  R., {Symonds}, K., {Rengel}, M., {Renotte}, E., {Ade}, P., {Affer}, L.,
  {Alard}, C., {Allard}, N., {Altieri}, F., {Andr{\'e}}, Y., {Arena}, C.,
  {Argyriou}, I., {Aylward}, A., {Baccani}, C., {Bakos}, G., {Banaszkiewicz},
  M., {Barlow}, M., {Batista}, V., {Bellucci}, G., {Benatti}, S., {Bernardi},
  P., {B{\'e}zard}, B., {Blecka}, M., {Bolmont}, E., {Bonfond}, B., {Bonito},
  R., {Bonomo}, A.~S., {Brucato}, J.~R., {Brun}, A.~S., {Bryson}, I., {Bujwan},
  W., {Casewell}, S., {Charnay}, B., {Pestellini}, C.~C., {Chen}, G.,
  {Ciaravella}, A., {Claudi}, R., {Cl{\'e}dassou}, R., {Damasso}, M.,
  {Damiano}, M., {Danielski}, C., {Deroo}, P., {Di Giorgio}, A.~M., {Dominik},
  C., {Doublier}, V., {Doyle}, S., {Doyon}, R., {Drummond}, B., {Duong}, B.,
  {Eales}, S., {Edwards}, B., {Farina}, M., {Flaccomio}, E., {Fletcher}, L.,
  {Forget}, F., {Fossey}, S., {Fr{\"a}nz}, M., {Fujii}, Y.,
  {Garc{\'\i}a-Piquer}, {\'A}., {Gear}, W., {Geoffray}, H., {G{\'e}rard},
  J.~C., {Gesa}, L., {Gomez}, H., {Graczyk}, R., {Griffith}, C., {Grodent}, D.,
  {Guarcello}, M.~G., {Gustin}, J., {Hamano}, K., {Hargrave}, P., {Hello}, Y.,
  {Heng}, K., {Herrero}, E., {Hornstrup}, A., {Hubert}, B., {Ida}, S., {Ikoma},
  M., {Iro}, N., {Irwin}, P., {Jarchow}, C., {Jaubert}, J., {Jones}, H.,
  {Julien}, Q., {Kameda}, S., {Kerschbaum}, F., {Kervella}, P., {Koskinen}, T.,
  {Krijger}, M., {Krupp}, N., {Lafarga}, M., {Landini}, F., {Lellouch}, E.,
  {Leto}, G., {Luntzer}, A., {Rank-L{\"u}ftinger}, T., {Maggio}, A.,
  {Maldonado}, J., {Maillard}, J.-P., {Mall}, U., {Marquette}, J.-B., {Mathis},
  S., {Maxted}, P., {Matsuo}, T., {Medvedev}, A., {Miguel}, Y., {Minier}, V.,
  {Morello}, G., {Mura}, A., {Narita}, N., {Nascimbeni}, V., {Nguyen Tong}, N.,
  {Noce}, V., {Oliva}, F., {Palle}, E., {Palmer}, P., {Pancrazzi}, M.,
  {Papageorgiou}, A., {Parmentier}, V., {Perger}, M., {Petralia}, A.,
  {Pezzuto}, S., {Pierrehumbert}, R., {Pillitteri}, I., {Piotto}, G., {Pisano},
  G., {Prisinzano}, L., {Radioti}, A., {R{\'e}ess}, J.-M., {Rezac}, L.,
  {Rocchetto}, M., {Rosich}, A., {Sanna}, N., {Santerne}, A., {Savini}, G.,
  {Scandariato}, G., {Sicardy}, B., {Sierra}, C., {Sindoni}, G., {Skup}, K.,
  {Snellen}, I., {Sobiecki}, M., {Soret}, L., {Sozzetti}, A., {Stiepen}, A.,
  {Strugarek}, A., {Taylor}, J., {Taylor}, W., {Terenzi}, L., {Tessenyi}, M.,
  {Tsiaras}, A., {Tucker}, C., {Valencia}, D., {Vasisht}, G., {Vazan}, A.,
  {Vilardell}, F., {Vinatier}, S., {Viti}, S., {Waters}, R., {Wawer}, P.,
  {Wawrzaszek}, A., {Whitworth}, A., {Yung}, Y.~L., {Yurchenko}, S.~N.,
  {Osorio}, M. R.~Z., {Zellem}, R., {Zingales}, T., \& {Zwart}, F. 2018,
  Experimental Astronomy, 46, 135

\bibitem[{{Venot} {et~al.}(2020){Venot}, {Cavali{\'e}}, {Bounaceur},
  {Tremblin}, {Brouillard}, \& {Lhoussaine Ben Brahim}}]{Venot20}
{Venot}, O., {Cavali{\'e}}, T., {Bounaceur}, R., {Tremblin}, P., {Brouillard},
  L., \& {Lhoussaine Ben Brahim}, R. 2020, \aap, 634, A78

\bibitem[{{Welbanks} {et~al.}(2019){Welbanks}, {Madhusudhan}, {Allard},
  {Hubeny}, {Spiegelman}, \& {Leininger}}]{Welbanks19}
{Welbanks}, L., {Madhusudhan}, N., {Allard}, N.~F., {Hubeny}, I., {Spiegelman},
  F., \& {Leininger}, T. 2019, \apjl, 887, L20

\bibitem[{{Wong} {et~al.}(2020){Wong}, {Shporer}, {Kitzmann}, {Morris}, {Heng},
  {Hoeijmakers}, {Demory}, {Ahlers}, {Mansfield}, {Bean}, {Daylan},
  {Fetherolf}, {Rodriguez}, {Benneke}, {Ricker}, {Latham}, {Vanderspek},
  {Seager}, {Winn}, {Jenkins}, {Burke}, {Christiansen}, {Essack}, {Rose},
  {Smith}, {Tenenbaum}, \& {Yahalomi}}]{Wong:2019aa}
{Wong}, I., {Shporer}, A., {Kitzmann}, D., {Morris}, B.~M., {Heng}, K.,
  {Hoeijmakers}, H.~J., {Demory}, B.-O., {Ahlers}, J.~P., {Mansfield}, M.,
  {Bean}, J.~L., {Daylan}, T., {Fetherolf}, T., {Rodriguez}, J.~E., {Benneke},
  B., {Ricker}, G.~R., {Latham}, D.~W., {Vanderspek}, R., {Seager}, S., {Winn},
  J.~N., {Jenkins}, J.~M., {Burke}, C.~J., {Christiansen}, J.~L., {Essack}, Z.,
  {Rose}, M.~E., {Smith}, J.~C., {Tenenbaum}, P., \& {Yahalomi}, D. 2020, \aj,
  160, 88

\bibitem[{{Zahnle} {et~al.}(2009){Zahnle}, {Marley}, {Freedman}, {Lodders}, \&
  {Fortney}}]{Zahnle09}
{Zahnle}, K., {Marley}, M.~S., {Freedman}, R.~S., {Lodders}, K., \& {Fortney},
  J.~J. 2009, \apjl, 701, L20

\bibitem[{Zellem {et~al.}(2019)Zellem, Swain, Cowan, \& {et
  al.}}]{Zellem:2019aa}
Zellem, R., Swain, M., Cowan, N., \& {et al.} 2019, Publications of the
  Astronomical Society of the Pacific, 131, 094401

\bibitem[{{Zhang} \& {Showman}(2018)}]{Zhang18b}
{Zhang}, X. \& {Showman}, A.~P. 2018, \apj, 866, 2

\end{thebibliography}

\end{document}